\documentclass[aps,prb,superscriptaddress,showpacs,twocolumn]{revtex4}
\usepackage{amsmath,amsfonts,bm,graphicx}

\begin{document}

\title{Influence of confining potentials on the exchange coupling in double quantum dots}
\author{Jesper Goor Pedersen}
\affiliation{DTU Fotonik, Department of Photonics Engineering, Technical University of Denmark, Building 343, DK-2800 Kongens Lyngby, Denmark}
\author{Christian Flindt}
\affiliation{Department of Physics, Harvard University, 17 Oxford Street, Cambridge, MA 02138, USA}
\author{Antti-Pekka Jauho}
\affiliation{DTU Nanotech, Department of Micro and Nanotechnology, Technical University of Denmark, Building 345east,
             2800 Kongens Lyngby, Denmark}
\affiliation{Aalto University, Department of Applied Physics,  P.\
O.\ Box 11100, FI-00076 AALTO, Finland}
\author{Niels Asger Mortensen}
\affiliation{DTU Fotonik, Department of Photonics Engineering, Technical University of Denmark, Building 343, DK-2800 Kongens Lyngby, Denmark}

\date{\today}

\begin{abstract}
We report simple expressions for the exchange coupling in double quantum dots calculated within the Heitler--London and the Hund--Mulliken approximations using four different confining potentials. At large interdot distances and at large magnetic fields the exchange coupling does not depend significantly on the details of the potentials. In contrast, at low fields and short distances different behaviors of the exchange coupling can be attributed to particular features of the potentials. Our results may be useful as guidelines in numerical studies and in the modeling of experiments.
\end{abstract}

\pacs{73.21.La, 75.30.Et}


\maketitle
\emph{Introduction}.--- The exchange coupling between electron spins in tunnel coupled quantum dots constitutes a key element in proposals for implementing quantum information processing in the solid state.\cite{Loss1998,Burkard1999} The exchange coupling splits the singlet and triplet spin states, depending on the confining potential and the applied magnetic field, thereby enabling electrical (or magnetic) control of the exchange coupling as demonstrated in recent experiments.\cite{Petta2005} The ability to control the exchange coupling with external fields, however, also makes the exchange coupling susceptible to electromagnetic fluctuations in the environment. A current trend is thus to search for ``sweet spots'' in parameter space,\cite{Coish2005,Hu2006,Stopa2008,Li2010} i.e., local maxima of the exchange coupling as function of external fields, where the exchange coupling to first order is insensitive to fluctuations.

Calculations of the exchange coupling can be approached with a variety of analytic and numerical methods. These include several analytic approximations\cite{Burkard1999} and numerical schemes such as exact diagonalization,\cite{Szafran2004, Helle2005, Zhang2006, Pedersen2007} path integral Monte Carlo simulations,\cite{Zhang2008} and configuration interaction calculations combined with density functional theory.\cite{Stopa2008} Numerical methods allow for calculations of the exchange coupling with high precision. However, in order to gain an understanding of the dependence of the exchange coupling on different parameters or as guidelines in the search for sweet spots,\cite{Coish2005,Hu2006,Stopa2008,Li2010} closed-form analytic expressions can be very useful.

In this Brief Report, we present simple analytic expressions for the exchange coupling of a double quantum dot obtained within the Heitler-London and Hund-Mulliken approximations\cite{Burkard1999} for four different confining potentials, one of which is new. We provide a comparative study of the exchange coupling calculated analytically for the four potentials as functions of interdot distance and magnetic field. In particular, we identify certain properties that are only weakly dependent on the choice of potential and discuss other features that, in contrast, can be associated with particular details of the potentials.

\begin{figure}
\begin{center}
\includegraphics[width=.9\linewidth]{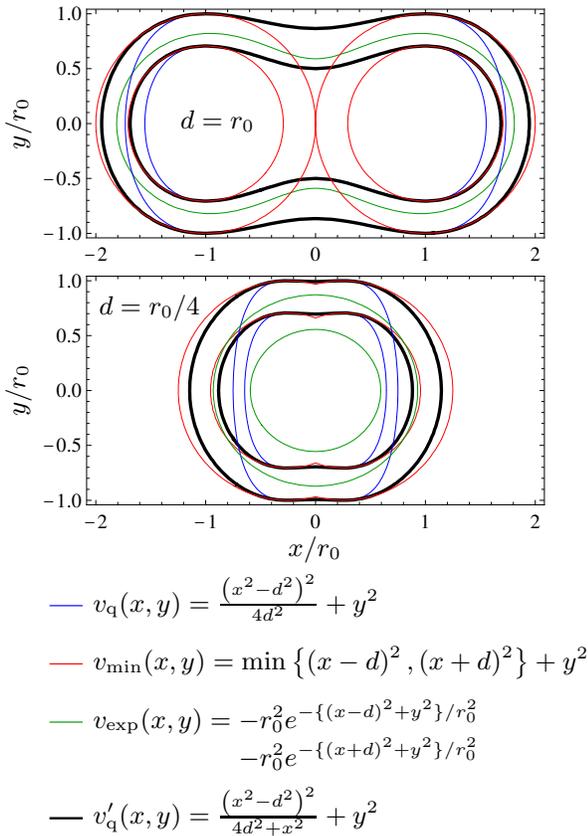}
\caption{(Color online) Contour plots of the model potentials at two different interdot distances.
The potentials are $V_i(x,y)\equiv (m\omega_0^2/2)v_i(x,y)$. The contours correspond
to $v(x,y)=r_0^2/4$ and $v(x,y)=r_0^2/2$. The potential $v_{\rm exp}$ has been shifted such that $v_{\rm exp}(\pm d,0)=0$.}
\label{fig:potentials}
\end{center}
\end{figure}

\emph{Model}.--- We consider two electrons confined by a double quantum dot in two dimensions in a perpendicular magnetic field. The two-electron Hamiltonian is
\begin{equation}
H(\mathbf{r}_1,\mathbf{r}_2) = h(\mathbf{r}_1)+
h(\mathbf{r}_2)+C(|\mathbf{r}_1-\mathbf{r}_2|),
\label{eq_hamiltonian}
\end{equation}
where
\begin{equation}
C(|\mathbf{r}_1-\mathbf{r}_2|)=\frac{e^2}{4\pi\varepsilon_r\varepsilon_0|\mathbf{r}_1-\mathbf{r}_2|}
\end{equation}
is the Coulomb interaction and the single-particle Hamiltonian in the effective-mass approximation is
\begin{equation}
h(\mathbf{r})=
\frac{1}{2m}\left[\mathbf{p}+e\mathbf{A}(\mathbf{r})\right]^2+V(\mathbf{r}),\,\,\ \mathbf{r}=(x,y).
\end{equation}
The confining potential is denoted as $V(\mathbf{r})$, $m$ is the effective electron mass, and
$\mathbf{A}(\mathbf{r})=B_z(-y,x)/2$ the magnetic vector potential. The Zeeman splitting does not affect the exchange coupling and has been omitted above. We use parameters typical of GaAs and take $m=0.067m_e$ and $\varepsilon_r=12.9$. The exchange coupling $J_V(B_z)=E_T-E_S$ is the difference between the spin triplet and spin singlet orbital ground states, respectively.

The four potentials considered in this work are defined in Fig.\ \ref{fig:potentials}
with $r_0=\sqrt{\hbar/m\omega_0}$ being the oscillator length, $\hbar\omega_0$ the confinement energy, and $2d$ the center to center distance between the dots. The first three potentials have previously been considered in the literature,\cite{Burkard1999,Szafran2004,Calderon2006,Helle2005,Hu2006,Zhang2006,Pedersen2007,Zhang2008,Li2010} while the last is new. The following approximations take as starting point the uncoupled dots at large distances, $d\gg r_0$. For the left/right dot centered at $\mathbf{r}_{L/R}=(\mp d,0)$, the ground state can be written\footnote{The ground state of  $V_{\mathrm{exp}}$ at large interdot distances cannot be solved analytically, but we have checked that the Fock--Darwin state is still a good approximation to the ground state found by numerical diagonalization.} as
$\varphi_{\mp d}\left(x,y\right)=\left<\mathbf{r}|L/R\right>=e^{\pm iyd/2l_B^2}\varphi\left(x\mp d, y\right)$ in terms of the Fock--Darwin ground state
$\varphi\left(x,y\right)=\sqrt{\frac{m\omega}{\pi \hbar}}e^{-m\omega\left(x^2+y^2\right)/2\hbar}$. Here, the magnetic length is $l_B=\sqrt{\hbar c/eB_z}$ and
$\omega=b\omega_0$, where $b=\sqrt{1+\omega_L^2/\omega_0^2}$ is the magnetic compression factor and
$\omega_L=eB_z/2mc$ the Larmor frequency. Additionally, we shall need the overlap $S\equiv\left<L|R\right>=e^{-d^2(2b-1/b)}$.

\emph{Heitler--London}.--- Within this
approximation the exchange coupling is estimated as
\begin{equation}
J^{\left(\mathrm{HL}\right)}=\left<-\right|H\left|-\right>-\left<+\right|H\left|+\right>,
\end{equation}
where $\left|\pm\right>=\left(\left|LR\right>\pm\left|RL\right>\right)/
\sqrt{2\left(1\pm S^2\right)}$. The exchange coupling is composed of contributions $J^{\left(\mathrm{HL}\right)}=
J^{\left(\mathrm{HL}\right)}_C+J^{\left(\mathrm{HL}\right)}_h$
from the Coulomb interaction and from the single-particle Hamiltonians, respectively. The first term $J^{\left(\mathrm{HL}\right)}_C$ does not depend on the potential and is given by Eq.~(7) in Ref.\ \onlinecite{Burkard1999}.  The second term $J^{\left(\mathrm{HL}\right)}_h$ is listed in Table \ref{table:JVcontributions} for each of the four potentials. The result for $V_{\mathrm{q}}$ has previously been reported in Ref.\ \onlinecite{Burkard1999}. The Heitler--London approximation is typically reliable when the ratio
$\frac{e^2}{4\pi\epsilon_r\epsilon_0r_0}/\hbar\omega_0$ is small.\cite{Calderon2006}

\begin{table*}
\begin{center}
\begin{tabular}{|l|c|}
\hline
\hline
\multicolumn{2}{|c|}{$J_h^{(\mathrm{HL})}/(\hbar\omega_0)$} \\
\hline
$V_\mathrm{q}$ & $\frac{2S^2}{1-S^4}
\frac{3}{4b}\left(1+bd^2\right)$ \\
\hline
$V_\mathrm{min}$ & $\frac{2S^2}{1-S^4}
\left[ \frac{2d}{\sqrt{b\pi}}\{1-e^{-bd^2}\}+2d^2\mathrm{erfc}(d\sqrt{b})
\right]$ \\
\hline
$V_\mathrm{exp}$ & $\frac{2S^2}{1-S^4}
\left[ \frac{d^2}{b^2} -\frac{b}{1+b}\left\{1+e^{-4bd^2/(1+b)}-2e^{-d^2/(b^2+b)}\right\}\right]$ \\
\hline
$V_\mathrm{q}'$ & $\frac{2S^2}{1-S^4}
\left[2d^2-\frac{25}{2}\sqrt{b\pi}d^3\left(
e^{4bd^2}\mathrm{erfc}\left(2\sqrt{b}d\right)-
\mathrm{Re}\left\{e^{(3+4i)bd^2}\mathrm{erfc}\left((2+i)\sqrt{b}d\right)\right\}\right)\right]$\\
\hline
\hline
\end{tabular}
\caption{The contribution $J^{\left(\mathrm{HL}\right)}_h$ to the exchange coupling within the Heitler-London approximation for the four potentials. The complementary error function is denoted as $\mathrm{erfc}(x)$.} \label{table:JVcontributions}
\end{center}
\end{table*}

\emph{Hund--Mulliken}.--- The Heitler--London approximation considers only the
singly occupied singlet and triplet states. When the tunnel coupling between the quantum dots becomes large, the doubly occupied spin singlet states should also be taken into account. The exchange coupling is then obtained by diagonalizing the Hamiltonian in the Hilbert space spanned by
$\Psi^D_{\pm d}(\mathbf{r}_1,\mathbf{r}_2)=\Phi_{\pm d}(\mathbf{r}_1)\Phi_{\pm d}(\mathbf{r}_2)$
and
$\Psi^S_{\pm}(\mathbf{r}_1,\mathbf{r}_2)=
[\Phi_{+d}(\mathbf{r}_1)\Phi_{-d}(\mathbf{r}_2) \pm  \Phi_{-d}(\mathbf{r}_1)\Phi_{+d}(\mathbf{r}_2)]/\sqrt{2}$,
where $\Phi_{\pm d}$ are the orthonormalized single-particle states
$\Phi_{\pm d} = (\varphi_{\pm d}-g\varphi_{\mp d})/\sqrt{1-2Sg+g^2}$ with
$g=(1-\sqrt{1-S^2})/S$. The exchange coupling is now estimated as\cite{Burkard1999}
\begin{equation}
J^{\left(\mathrm{HM}\right)}=V-U_r/2+\frac{1}{2}\sqrt{U_r^2+16t^2_r}.
\label{eq:HM}
\end{equation}
Here, $U_r$ and $t_r=t-w = -\left< \Phi_{\pm d} \right|h\left| \Phi_{\mp d}\right>
-\left<\Psi_+^S\right|C\left|\Psi_{\pm d}^D\right>/\sqrt{2}$ are the renormalized on-site Coulomb interaction and tunnel coupling, respectively, and $V$ (not to be confused with the confining potential)
is the difference in Coulomb energy between the singly occupied singlet and
triplet states. We find that the bare tunnel coupling can be written as $t=J_h^{(\mathrm{HL})}(1+S^2)/4S$, where $J_h^{(\mathrm{HL})}$ is given in Table~\ref{table:JVcontributions}, and $U_r$, $w$, and $V$ can be found in Appendix A of Ref.~\onlinecite{Burkard1999}.

\begin{figure}
\begin{center}
\includegraphics[width=0.95\linewidth]{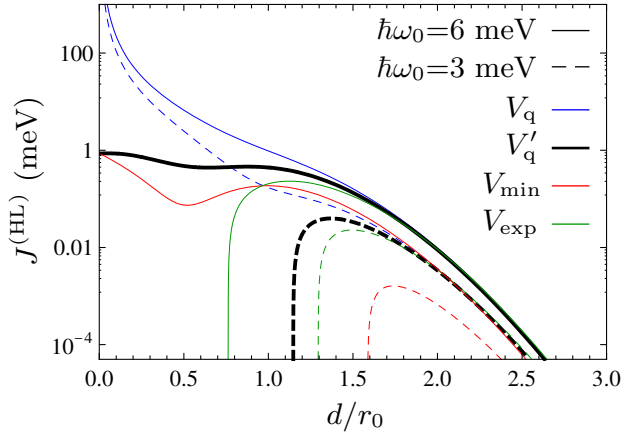}
\caption{(Color online) Exchange coupling in the Heitler--London
approximation as a function of the interdot
distance. Two different confinement strengths $\hbar\omega_0$ are used.
}
\label{fig:HL}
\end{center}
\end{figure}

\emph{Results}.--- In Fig.~\ref{fig:HL} we show the exchange coupling $J^{\left(\mathrm{HL}\right)}$ in the
Heitler--London approximation at zero magnetic field as function of the interdot distance. We note that the exchange coupling at zero magnetic field always must be non-negative. At large distances, $d>r_0$, the potentials separate into two isolated quantum dots with qualitatively similar behavior, showing a decay of the exchange coupling with increasing interdot distance. The potential $V_{\mathrm{min}}$ results in a slightly lower exchange coupling compared to the other potentials at a given large interdot distance. At short distances, in contrast, the results depend strongly on the particular choice of potential. For both confinement energies, the potential $V_\mathrm{q}$ yields non-negative results. However, at short distances, $d\ll r_0$, the exchange coupling for this potential diverges. In contrast, the three other potentials yield (un-physical) negative exchange couplings (near the abrupt decreases of the exchange coupling in the logarithmic plots) within the Heitler--London approximation at short distances and low confinement energies. This behavior is well-known for the Heitler-London approximation.\cite{Calderon2006,Pedersen2007} Eventually, the Heitler--London approximation also breaks down for the potential $V_\mathrm{q}$ and predicts a negative exchange coupling, however, only at lower confinement energies (not shown). At large confinement energies, the potentials $V_\mathrm{min}$ and $V_\mathrm{q}'$ give non-negative, finite values of the exchange coupling at short distances.

\begin{figure}
\begin{center}
\includegraphics[width=0.95\linewidth]{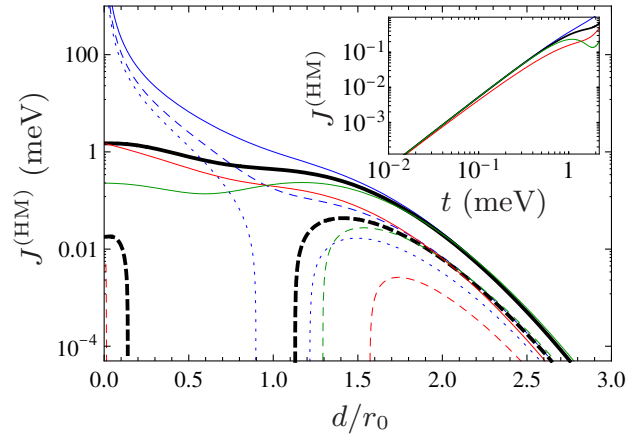}
\caption{(Color online) Exchange coupling in the Hund--Mulliken
approximation as function of the interdot
distance  (see Fig.~\ref{fig:HL} for legend). Two different confinement strengths are used. We also show results for $V_\mathrm{q}$ with $\hbar\omega_0=2$ meV (dotted  line). In the inset the exchange coupling as function of the bare tunnel coupling $t$ illustrates the general $t^2$-dependence at large interdot distances.
}
\label{fig:HM}
\end{center}
\end{figure}

In Fig.~\ref{fig:HM} we show results for the exchange coupling in the Hund--Mulliken approximation at zero magnetic field. At large distances, $d\gg r_0$, the exchange coupling is again similar for the four potentials.  This behavior is further corroborated by the inset of Fig.~\ref{fig:HM}, showing the exchange coupling as function of the bare tunnel coupling  $t$. At large distances, where the tunnel coupling is small, a clear $t^2$-dependence is found as expected in the Hubbard model picture.
At short distances, $d\lesssim  r_0$, and at low confinement energies, $\hbar\omega_0=3$ meV, only the potential $V_{\mathrm{q}}$ gives non-negative values of the exchange coupling. However, at lower confinement energies, $\hbar\omega_0=2$ meV, the Hund--Mulliken approximation also breaks down for this potential around $d=r_0$ (dotted line in Fig.~\ref{fig:HM}). At higher confinement energies, $\hbar\omega_0=6$ meV, all potentials yield positive values of the exchange coupling. Also in the Hund--Mulliken approximation, the potential $V_{\mathrm{q}}$ results in a diverging exchange coupling at low interdot distances.

\begin{figure}
\begin{center}
\includegraphics[width=0.90\linewidth]{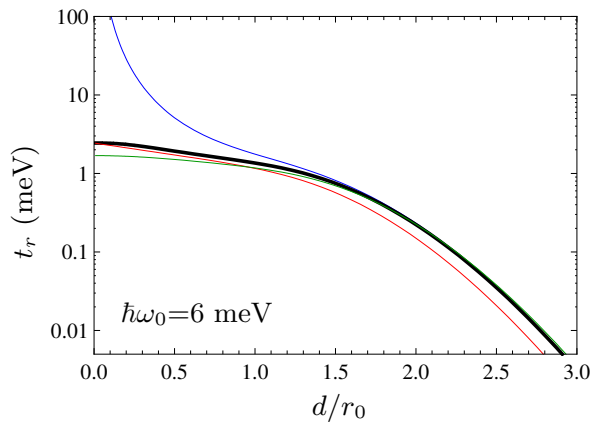}
\caption{(Color online) Renormalized tunnel coupling as function
of interdot distance (see Fig.~\ref{fig:HL} for legend).
}
\label{fig:tr}
\end{center}
\end{figure}

In order to explain the observed trends, we consider Eq.\ (\ref{eq:HM}) for the exchange coupling in the Hund--Mulliken approximation. At large confinement energies, the renormalized tunnel coupling may ensure a well-behaved limit at short interdot distances. At the same time, a too small renormalized tunnel coupling $t_r$ can result in negative values of the exchange coupling. In Fig.\ \ref{fig:tr}, we show the renormalized tunnel coupling $t_r$ as function of the interdot distance. At large interdot distances, the renormalized tunnel coupling is similar for the four potentials, although somewhat smaller for $V_{\mathrm{min}}$. The smaller tunnel coupling is due to the height of the barrier separating the two dots being larger for $V_{\mathrm{min}}$ compared to the other potentials. This explains the results for the exchange coupling at large interdot distances within the Hund--Mulliken approximation shown in Fig.~\ref{fig:HM}. At short distances, the potentials $V_\mathrm{exp}$, $V_\mathrm{min}$, and $V_\mathrm{q}'$ yield qualitatively identical results with well-behaved limits for the renormalized tunnel coupling and, consequently, also well-behaved values for the exchange couplings in Fig.~\ref{fig:HM}. In contrast, the renormalized tunnel coupling and the exchange energy diverge for $V_\mathrm{q}$ at short interdot distances. The large renormalized tunnel coupling at intermediate distances ensures positive values of the exchange coupling in the Hund--Mulliken approximations, but eventually leads to a diverging exchange energy at short distances.

We attribute the divergence of the renormalized tunnel coupling for $V_\mathrm{q}$ to the particular behavior of the potential at short interdot distances. Unlike the three other potentials, the potential $V_\mathrm{q}$ does not simplify to a single-dot potential at $d=0$ (the potential $V_\mathrm{exp}$ already collapses into a single wide dot at $d\lesssim 3r_0/4$). In fact, it contains a diverging term of the form $x^4/d^2$. To remedy this problem, we introduced the potential $V_\mathrm{q}'$ with the modified denominator $4d^2\rightarrow 4d^2+x^2$. At short interdot distances we then have  $V_\mathrm{q}'(x,0)\sim (x^2-d^2)^2/(4d^2+x^2)\rightarrow x^2$ corresponding to a single harmonic potential. At large distances, $d\gg r_0$, the potential $V_\mathrm{q}'$ is similar to $V_\mathrm{q}$, while the modified denominator ensures well-behaved limits for the exchange coupling and the renormalized tunnel coupling as illustrated in Figs.\ \ref{fig:HM} and \ref{fig:tr}.

\begin{figure}
\begin{center}
\includegraphics[width=0.90\linewidth]{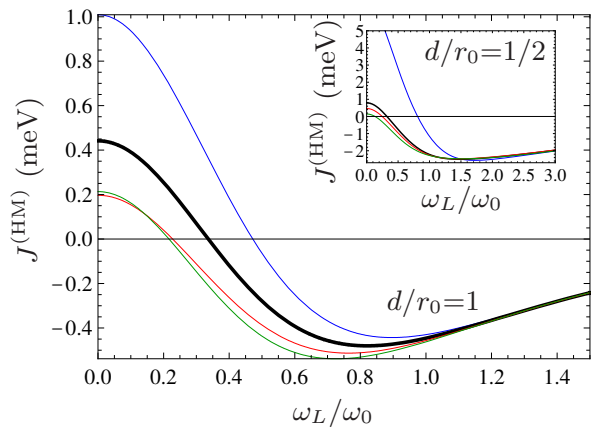}
\caption{(Color online) Exchange coupling in the Hund--Mulliken
approximation as function of the magnetic field (see Fig.~\ref{fig:HL} for legend).
Results are shown for $\hbar\omega_0=6$~meV and two different interdot distances.
}
\label{fig:HMvsB}
\end{center}
\end{figure}

Finally, we turn to the magnetic field dependence of the exchange coupling.  The Hund--Mulliken approximation is typically more reliable than the Heitler--London approximation in predicting the magnetic field dependence. In Fig.~\ref{fig:HMvsB} we consequently show Hund--Mulliken predications of the exchange coupling as function of the applied magnetic field. At large magnetic fields, the exchange coupling is similar for the four potentials, whereas different behaviors are seen at low fields, corresponding to the differences seen at zero magnetic field in Fig.~\ref{fig:HM}. At large fields, magnetic compression suppresses the tunnel coupling and we recover the $J\propto t^2$ dependence also seen at large interdot  distances with zero magnetic field, as illustrated in the inset of Fig.~\ref{fig:HM}. At short interdot distances and small magnetic field, the exchange coupling for $V_\mathrm{q}$ again blows up due to the diverging term in the potential. In contrast, the behavior of the exchange coupling corresponding to $V_\mathrm{q}'$ is again well-behaved at small fields as seen in the inset of Fig.~\ref{fig:HMvsB}.

\emph{Conclusions}.--- We have studied the exchange coupling between electron spins in double quantum dots within the Heitler--London and Hund--Mulliken approximations using four different confining potentials. At large interdot distances and at high magnetic fields the exchange coupling is only weakly sensitive to the details of the potentials. In contrast, at short interdot distances the exchange coupling depends on the choice of potential. At short interdot distances and low magnetic fields, the potential $V_\mathrm{q}$ yields a diverging exchange coupling. We have slightly modified this potential in order to remedy this problem. The simple expressions for the exchange coupling presented in this work may by useful as guidelines in numerical studies and in the modeling of experimental setups.

\emph{Acknowledgements}.--- The work by CF was supported by the Villum Kann Rasmussen Foundation. APJ is
grateful to the FiDiPro program of the Finnish Academy.

\end{document}